\input{psfig}

\documentstyle[prl,aps,twocolumn]{revtex} % title2.tex
\catcode`\@=11

\def\maketitle2{\par % Uses \twocolumn[\@maketitle2].
\begingroup
\let\cite\@bylinecite
\def\thefootnote{\fnsymbol{footnote}}%
\twocolumn[\@maketitle2\vskip2pc]%
\thispagestyle{plain}\@thanks
\endgroup
\def\thefootnote{\arabic{footnote}}%
\setcounter{footnote}{0}%
\let\maketitle2\relax \let\@maketitle2\relax
\let\@thanks\relax \let\@authoraddress\relax \let\@title\relax
\let\@date\relax \let\thanks\relax \let\@abstract\relax 
\let\@pacs\relax}

\def\abstract#1{\gdef\@abstract{{\par % Store abstract text. 
\bgroup
\ifdim\prevdepth=-1000pt \prevdepth0pt\fi
\hsize\columnwidth
\dimen0=-\prevdepth \advance\dimen0 by17.5pt \nointerlineskip
\small\vrule width 0pt height\dimen0 \relax}{~~}#1\egroup}}

\def\pacs#1{\gdef\@pacs{{\par % Store PACS numbers as \@pacs.
\bgroup
\hsize\columnwidth \parindent0pt
\ifdim\prevdepth=-1000pt \prevdepth0pt\fi
\dimen0=-\prevdepth \advance\dimen0 by20pt\nointerlineskip
\egroup} PACS numbers:~#1}}

\def\@maketitle2{% Puts \@abstract and \@pacs in a {list}.
\@preprint
\@title
\ifdim\prevdepth=-1000pt \prevdepth0pt\fi
\@authoraddress
\@date
\begin{list}{}{\leftmargin=0.10753\textwidth \rightmargin=\leftmargin
\itemsep=1pc\partopsep=-1pc}
\item\@abstract
\item\@pacs
\end{list}
}

\catcode`\@=12

\begin{document}

\title{\vspace*{-0.5cm} \hspace*{\fill}{\normalsize LA-UR-98-4184}
\\[1.5ex] 
Bose-Einstein Correlations and the Equation of State of
Nuclear Matter 
} 
\author{ 
B.R. Schlei${}^1$\thanks{E. Mail: schlei@LANL.gov}{\ }, 
D. Strottman${}^2$\thanks{E. Mail: dds@LANL.gov}{\ }, 
J.P. Sullivan${}^1$\thanks{E. Mail: sullivan@LANL.gov}{\ } and 
H.W. van Hecke${}^1$\thanks{E. Mail: hubert@LANL.gov}{\ } \\[1.5ex] 
{\it  ${}^1$ Physics Division, P-25,
Los Alamos National Laboratory, Los Alamos, NM 87545, USA\\ 
${}^2$ LANSCE Division and Theoretical Division, Los Alamos National
Laboratory, Los Alamos, NM 87545, USA 
} 
} 
\date{\today}

\abstract{  
Within a relativistic hydrodynamic framework, we use four
different  equations of state of nuclear matter to compare to
experimental spectra from CERN/SPS experiments NA44 and
NA49. Freeze-out hypersurfaces and Bose-Einstein  correlation
functions for identical pion pairs are discussed. We find that
two-pion Bose-Einstein interferometry measures the relationship
between  the temperature and the energy density in the equation of
state during the late hadronic stage of the fireball expansion. Little
sensitivity of the light-hadron data to a quark-gluon plasma
phase-transition is seen.
}  
\maketitle2 
\narrowtext

The equation of state (EOS) of nuclear matter at very high energy
densities remains unknown.  A possible new state of nuclear matter,
the quark-gluon plasma (QGP), may be formed within a very hot and
dense zone of nuclear matter - the fireball - during relativistic
heavy-ion collision experiments.  Many observables have been proposed
as a signature for the QGP, among them  measurements derived  from
intensity interferometry of identical hadrons, also known as
Bose-Einstein correlations (BEC). BEC functions are sensitive to the
space-time dynamics of the fireball, and therefore  should give clues
about the EOS, which governs the evolution of those fireballs.

Among the many studies on BEC (for a recent overview,  {\it cf.}, the
book by Weiner \cite{wei97}) only few  ({\it cf}.,  e.g., Rischke {\it
et al.} \cite{ris96}) address the role  of the EOS of nuclear matter
in the intensity interferometry of  identical hadrons.  It is the
purpose of this paper to further investigate the interplay between the
EOS and BEC. In the following, we shall consider a framework of
analysis which is based on relativistic hydrodynamics, since this
approach allows for an explicit use of an EOS.

We shall use the simulation code HYLANDER-C \cite{brs97_2} with
various  equations of state, and Cooper-Frye \cite{coo75} freeze-out.
The results of these hydrodynamical (one-fluid-type) calculations
\cite{brs98} will be compared to experimental single particle momentum
distributions from 158$A$ GeV Pb+Pb collisions, measured by the NA44
\cite{nxu96,bea96} and NA49 \cite{jon96} Collaborations.  After a
discussion of the space-time features of the particular fireballs we
shall compare BEC of identical pion pairs \cite{new} to experimental
correlation functions, as measured by NA44 \cite{bea98}.

The first equation of state \cite{red86}, EOS-I, which we use in this
study
 exhibits a phase transition to a quark-gluon plasma at a critical
temperature $T_c$ = 200 $MeV$ with a critical energy density
$\epsilon_c$ = 3.2 $GeV/fm^3$.  The second equation of state
\cite{hun95}, EOS-II, is a lattice QCD-based EOS (as is EOS-I) which
has recently become very popular in the field of relativistic
heavy-ion physics.  This equation of state includes a phase transition
to a quark-qluon plasma at $T_c$ = 160 $MeV$ with a critical energy
density $\epsilon_c$ $\approx$ 1.5 $GeV/fm^3$. The third equation of
state, EOS-III, has been extracted from the microscopic transport
model RQMD \cite{sor97} under the assumption of complete
thermalization, and does {\it not} include a transition to a QGP.   We
obtain a fourth equation of state, EOS-Ib, by changing the
relationship between $\epsilon_c$ and $T_c$ in EOS-I to $T_c
(\epsilon_c = 1.35\: GeV/fm^3 )$ = 200 $MeV$.

In Fig. 1 the four equations of state are shown in two different
representations. The particular parametrizations for the EOS  which
have been used here are described in Ref. \cite{brs98} in more detail.

In the following, we shall discuss five scenarios: we compare four
calculations using EOS-I, EOS-II, EOS-III, and EOS-Ib, all at fixed
freeze-out temperature $T_f$ = 139 $MeV$, and one calculation using
EOS-I for fixed  freeze-out temperature $T_f$ = 116 $MeV$.

It is possible to find initial distributions ({\it cf.} Table II in
Ref. \cite{brs98}  and Table I (here)) for the four equations of
state, such that one can reproduce the single inclusive momentum
spectra of 158 $AGeV$ Pb+Pb collisions. For a freeze-out temperature
$T_f$ = 139 $MeV$, the initial conditions and a  large number of final
single inclusive momentum distributions for various  hadron species
have been shown in Refs. \cite{brs97_2,brs98}  in comparison to the
data measured by the NA44 \cite{nxu96,bea96}  and NA49 \cite{jon96}
Collaborations. Those results  refer to calculations using EOS-I,
EOS-II, and EOS-III. Results showing the  fits of single inclusive
particle momentum spectra using EOS-Ib with  $T_f$ = 139  $MeV$ and
EOS-I with $T_f$ = 116 $MeV$ and $T_f$ = 139  $MeV$ are shown in
Fig. 2.

The magnitudes of the slopes in the transverse mass spectra have their
origin in the freezeout temperature and the transverse velocity fields
at freeze-out.  Because the effective EOS softness \cite{brs98} is
larger in the calculations using EOS-Ib with $T_f$ = 139 $MeV$  and
EOS-I with $T_f$ = 116 $MeV$ compared to the calculation using EOS-I
with  $T_f$ = 139 $MeV$, the values for the maximum transverse
velocity at  freeze-out, $v_\perp^{max}$, are correspondingly larger
({\it cf.} Table I) in the calculations using EOS-Ib with $T_f$ = 139
$MeV$ and EOS-I with  $T_f$ = 116 $MeV$ compared to the calculation
using EOS-I with $T_f$ = 139  $MeV$. But because in the calculation
using EOS-I with $T_f$ = 116 $MeV$ the  temperature is decreased while
increasing the value for $v_\perp^{max}$,  hardly any change is seen
in the slopes of the transverse mass spectra  compared to the
calculation using  EOS-I with $T_f$ = 139 $MeV$. A more  pronounced
change is seen only in the calculation using the harder EOS-Ib while
keeping $T_f$ unchanged.

It should be stressed that all calculations discussed so far (except
the calculation using EOS-II) result in single particle momentum
distributions that describe the data equally well. Although EOS-II was
found in the calculations of hadronic transverse  mass spectra to be
too soft ({\it cf.} Refs. \cite{brs97_2,brs98}), we shall use it here
also for the calculation of Bose-Einstein correlation functions.

Before we discuss BEC, we briefly discuss the thermal
evolution of the  various fireballs.

Fig. 3 shows the isothermes for the relativistic Pb+Pb fluids governed
by EOS-I and EOS-Ib ({\it cf.} also Ref. \cite{brs98}) until
freeze-out has been  reached. In Ref. \cite{brs98} it was shown that
the calculation using EOS-I with $T_f$ = 139 $MeV$ leads to a fireball
of a much shorter liftime than the calculations using EOS-II and
EOS-III with $T_f$ = 139 $MeV$. This behavior is caused by much
smaller freeze-out energy densities, $\epsilon_f$, in the calculations
using EOS-II and EOS-III compared to the calculation using EOS-I. We
have $\epsilon_f$ = 0.292 [0.126 (0.130)] $GeV/fm^3$ when using EOS-I
[EOS-II (EOS-III)]. A fluid that undergoes adiabatic expansion needs
more time to reach the smaller freeze-out energy densities.

Since EOS-II and EOS-III yield similar lifetimes of the fireball, we
attempt in the following to increase the lifetime of the system which
is governed by EOS-I. This can be achieved by (a) using a smaller
freeze-out temperature $T_f$ = 116 $MeV$, or (b) by hardening the EOS,
i.e., using EOS-Ib instead of EOS-I (without changing $T_f$ = 139
$MeV$).  For the latter two cases, we obtain $\epsilon_f$ = 0.127
$GeV/fm^3$ ({\it cf.}  Fig. 1 (c) and Ref. \cite{new}).

The lifetime of the fireball is reflected in Bose-Einstein
correlations of identical pion pairs \cite{brs97}. Using the
Bertsch-Pratt parametrization \cite{ber88}, a sensitive quantity is
the longitudinal  projection of the two-pion correlation function,
$C_2(q_{long})$, since it  reflects the effects of transverse
expansion \cite{brs92} as well as
 the contributions of resonance decay \cite{brs93} in BEC.

Fig. 4 shows data points taken by the NA44  Collaboration
\cite{bea98}, along with  projections of calculated BEC functions for
$\pi^+\pi^+$ pairs, using the same acceptance as the experiment.
Consistent with expectation, the calculations using EOS-II,
EOS-III, and EOS-Ib give similar lifetimes and therefore sufficiently
large longitudinally expanded fireballs (see Figs. 2,3). These
calculations result in a good reproduction of the pionic NA44 BEC
data. In addition, EOS-III and EOS-Ib  produce an excellent
description \cite{brs97_2,brs98,new} of hadronic single inclusive
momentum spectra.

It should be stressed here that a  freeze-out temperature $T_f$ = 139
$MeV$ was {\it adequate} to achieve this  agreement.

Also consistent with expectation is that the calculation using EOS-I
with  $T_f$ = 139 $MeV$ gave a  longitudinally expanded fireball 
\cite{brs98}  which was too small. On the contrary, it is initially 
surprising that a reduction of the freeze-out
temperature  to $T_f$ = 116 $MeV$ does not lead to a large enough
longitudinal extension of the  fireball in the calculation which uses
EOS-I.
The reason for this result is the following:  a freeze-out
temperature reduction leads to a larger lifetime of  the {\it direct}
fireball, but because of the lower temperature, the relative fraction
of heavy resonance decay contributions is reduced (by about 30\%), so
that the 'resonance halo' is  reduced in size. Hence, the {\it
apparent} fireball, which is a superposition of the direct (or
thermal) fireball and the resonance halo remains more or less
unchanged in size.

We note that the transverse projections of the correlation functions
$C_2(q_{out})$ and $C_2(q_{side})$ are also described reasonably well,
especially by those calculations which yield the larger {\it
longitudinal} freeze-out extension. The numbers in Fig. 4 can be
obtained from a  1-dimensional fit of the correlation functions with
\cite{brs98_2} \begin{eqnarray} C_2(\vec{k}_1,\vec{k}_2)&=&
1\:+\:\lambda(\vec{K})\cdot\exp [-\:q_l^2 R_l^2(\vec{K})\:-\:q_o^2
R_o^2(\vec{K})\: \nonumber\\ &&-\:q_s^2 R_s^2(\vec{K})\:
+2\:\rho_{ol}(\vec{K}) q_o q_l R_o(\vec{K}) R_l(\vec{K})] \:,
\nonumber\\ \label{eq:schlei} \end{eqnarray}

but they should not be taken too seriously.  In eq. (\ref{eq:schlei})
the $q_i$ ($i = l, s, o$) refer to the components of the momentum
difference $\vec{q} = \vec{k}_1 - \vec{k}_2$, and $\vec{K} = 
\textstyle{\frac{1}{2}} (\vec{k}_1 + \vec{k}_2$) is the average
pion pair momentum. Furthermore, $\lambda(\vec{K})$
is the momentum dependent incoherence factor which accounts for
reductions of the BEC due to long-lived resonances \cite{brs93} and
averaging due to phase-space, respectively.

In conclusion, by inspecting Fig. 1(b) we can see that only those
equations of state which go through point $C$ in Fig. 1 reproduce
the experimental  data on Bose-Einstein correlations fairly
well. Changing the freeze-out temperature shows hardly any
effect. Therefore, the measurements of Bose-Einstein correlations tell
us which relationship between temperature and energy density is
neccessary for a valid choice of an equation of state in the
calculations.  Unfortunately, from the above considerations it
must be noted that a two-particle BEC {\it used by itself} cannot be
used as a tool  to determine a possible phase-transition to a QGP,
because the BEC show little  sensitivity to the structure of the EOS.

\bigskip One of us, BRS, would like thank M. Gyulassy for instructive
discussions.  This work has been supported by the U.S. Department of
Energy.

\vspace*{-0.3cm} \begin{table} \caption{Properties of the fireballs.}
\begin{center} \begin{tabular}{p{4.5cm} c c c}
 & EOS-I & EOS-I & EOS-Ib \\ \hline \multicolumn{4}{c}{Initial
parameters}\\ Rel.  fraction of thermal energy in the central
fireball, $K_L$ & 0.55 & 0.60 & 0.60 \\ Longitudinal extension of the
central fireball, $\Delta$ $[fm]$ & 1.20 & 1.20 & 1.20 \\ Rapidity at
the edge of the central fireball, $y_\Delta$ & 1.00 & 1.05 & 1.00 \\
Rapidity at maximum of initial baryon $y$ distribution, $y_m$ & 0.80 &
0.85 & 0.85 \\ Width of initial baryon $y$ distribution, $\sigma$ &
0.32 & 0.32 & 0.32 \\ Freeze-out temperature, $T_f$ $[MeV]$ & 139 &
116 & 139 \\ \multicolumn{4}{c}{Output}\\ Max.  initial energy
density, $\epsilon_\Delta$ $[GeV/fm^3]$ & 15.3 & 16.8 & 16.8 \\ Max.
initial baryon density, $n_B^{max}$ $[fm^{-3}]$ & 3.93 & 4.11 & 3.90
\\ Rel.  fraction of baryons in central fireball, $f_{n_B}^\Delta$ &
0.71 & 0.78 & 0.73 \\ Freeze-out energy density, $\epsilon_f$
$[GeV/fm^3]$ & 0.292 & 0.127 & 0.127 \\ Max.  transverse velocity at
freeze-out, $v_\perp^{max}$ $[c]$ & 0.46 & 0.61 & 0.58 \\ Lifetime of
fireball, $t_{max}$ $[fm/c]$ & 13.1 & 20.2 & 20.5 \\ Lifetime of QGP,
$t_{QGP}$ $[fm/c]$ & 2.4  & 2.6  & 6.0 \\ \end{tabular} \end{center}
\end{table}

\vspace*{-6.5cm} \begin{figure} \begin{center}\mbox{ }
\psfig{figure=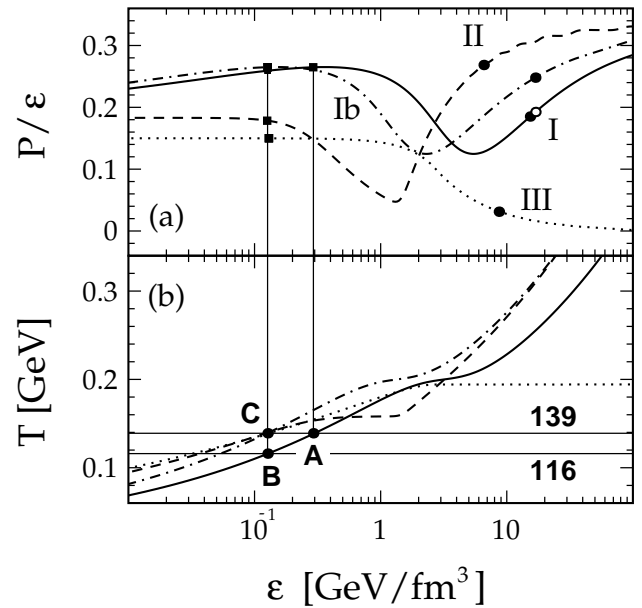,bbllx=3.0cm,bblly=3.0cm,%
bburx=20.0cm,bbury=27.7cm,width=9.5cm,clip=} \caption{Ratio of
pressure and energy density, $P/\epsilon$, and temperature, $T$, as
functions of $\epsilon$,  for the equations of state EOS-I (solid
lines), EOS-II (dashed lines),  EOS-III (dotted lines), and EOS Ib
(dashed-dotted lines), respectively.  The dots in plot (b) correspond
for each EOS to the starting values of $P/\epsilon$ with respect to
the achieved initial maximum energy density $\epsilon_\Delta$ at
transverse position $r_\perp = 0$, (dots correspond to $T_f = 139$ MeV
whereas the open circle corresponds to $T_f = 116$ MeV). The squares
indicate the final values of $P/\epsilon$  at breakup energy
densities, $\epsilon_f$. The dots $A$, $B$, $C$ in plot (c) indicate
the relationship between the temperature and the energy density  at
the late hadronic stage of the fireball expansion.}  \end{center}
\label{fg:fig1} \end{figure}

\newpage \begin{figure}[h] \begin{center}\mbox{ }
\psfig{file=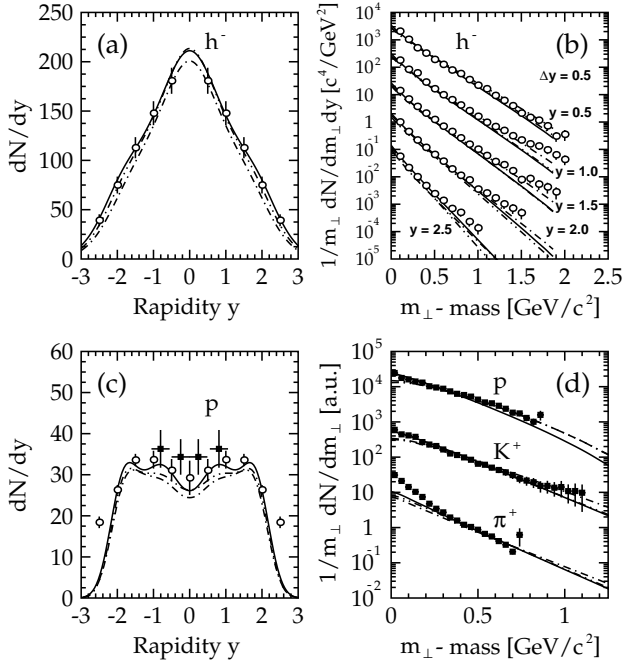,bbllx=1.1cm,bblly=3.0cm,%
bburx=21.0cm,bbury=28.0cm,width=8.2cm,clip=} \caption{ (a) Rapidity
spectra and (b) transverse mass spectra, $1/m_\perp dN/dm_\perp dy$,
of negative hadrons, $h^-$, (c) rapidity spectra of protons (without
contributions from $\Lambda^0$ decay) and (d) transverse mass spectra,
$1/m_\perp dN/dm_\perp$, of protons (including contributions from
$\Lambda^0$ decay), $p$, positive kaons, $K^+$, and positive pions,
$\pi^+$, respectively.  The solid (double dotted-dashed
[dashed-dotted]) lines indicate  the results of the calculations when
using equation of state EOS-I (EOS-I  [EOS-Ib]) with $T_f$ = 139 (116
[139]) $MeV$. The open circles represent preliminary data taken by the
NA49 Collaboration, whereas the filled squares represent final data
taken by the NA44 Collaboration.  } \end{center} \label{fg:fig2}
\end{figure}

\begin{figure} \vspace*{-2.0cm} \begin{center}\mbox{ }
\psfig{figure=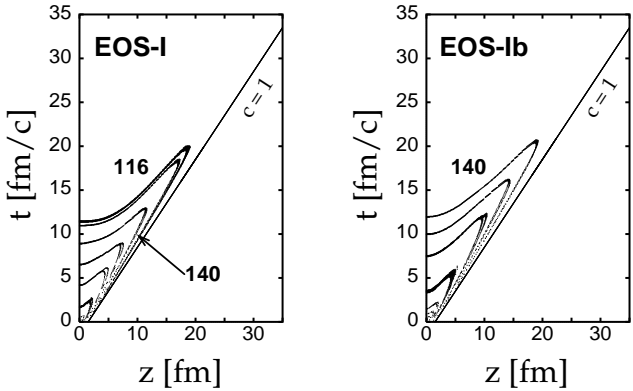,bbllx=1.5cm,bblly=1.5cm,%
bburx=21.0cm,bbury=14.0cm,width=9.3cm,clip=} \caption{ Isothermes for
the relativistic fluids governed by EOS-I, and EOS-Ib at $r_\perp =
0$, respectively. For EOS-I the lines (beginning with the most outer
curves) correspond to temperatures, $T$ = 116 $MeV$, 120 $MeV$,  140
$MeV$, 160 $MeV$ ... etc. For EOS-Ib, the outer lines correspond to a
temperature, $T$ = 140 $MeV$, and each successively smaller curve
represents a reduction in temperature by $\Delta T$ = 20 $MeV$.   The
lines $c=1$ represent the light cone.}  \end{center} \label{fg:fig3}
\end{figure}

\begin{figure} \vspace*{-1.8cm} \begin{center}\mbox{ }
\psfig{figure=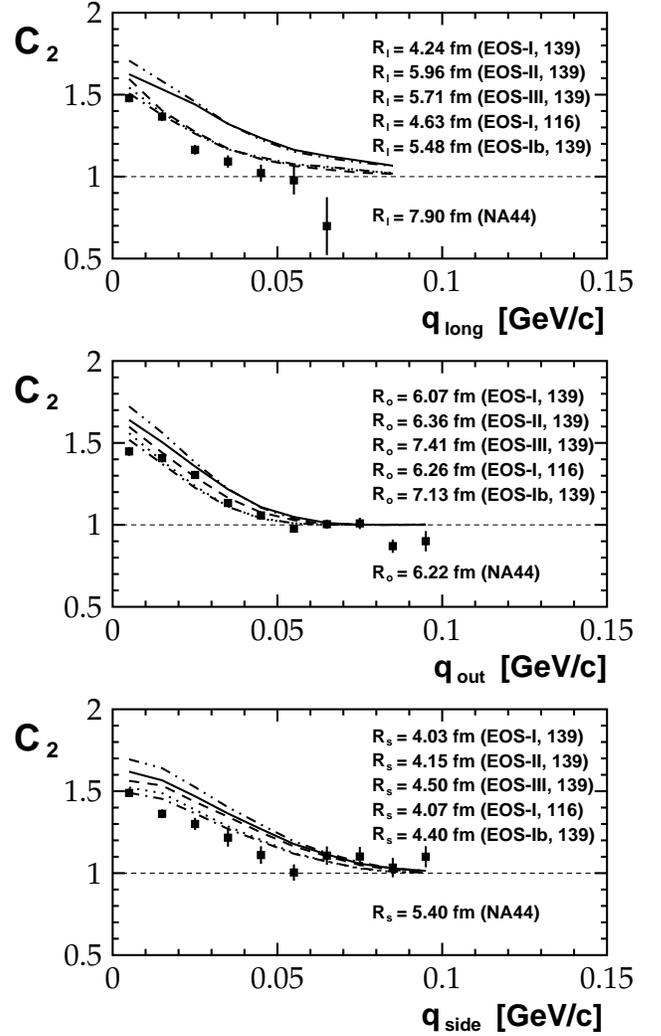,bbllx=2.2cm,bblly=0.5cm,%
bburx=19.0cm,bbury=29.7cm,width=9.2cm,clip=} \caption{Projections of
BEC functions for $\pi^+\pi^+$ pairs emerging from  the low $p_\perp$
(horizontal and vertical) acceptance setting of the NA44  detector
[8]. The data points are data taken by the NA44 Collaboration [8].
The solid [dashed [dotted [dashed-dotted]]] lines correspond  to the
calculations using EOS-I [EOS-II [EOS-III [EOS-Ib]]] with $T_f$ = 139
$MeV$, and the double dotted-dashed lines correspond to the
calculation using EOS-I with $T_f$ = 116 $MeV$. The values of the
inverse width  parameters, $R_i$ ($i=l,o,s$), have been obtained from
the Gaussian Bertsch-Pratt parametrization (see text).}  \end{center}
\label{fg:fig4} \end{figure}

\end{document}